\documentclass{arxivpreprint}

\usepackage{amsmath,amssymb,bm}
\usepackage{float}
\usepackage{enumitem}
\usepackage{ragged2e}
\graphicspath{{./figures/}}

\begin{document}
\justifying
\fancyhead[R]{G F Pinton}

\articletype{Paper}

\title{A reduced viscoelastic FDTD formulation for ultrasound-driven shear wave propagation in soft tissue}

\author{Gianmarco F Pinton$^{1,*}$\orcid{0000-0002-4896-1439}}

\affil{$^1$Lampe Joint Department of Biomedical Engineering, University of North Carolina at Chapel Hill and North Carolina State University, Chapel Hill, NC, USA}

\affil{$^*$Author to whom any correspondence should be addressed.}

\email{gia@email.unc.edu}

\keywords{shear wave elastography, FDTD, acoustic radiation force, viscoelasticity, Kelvin--Voigt}

\begin{abstract}
Ultrasound-driven shear wave propagation in soft tissue underlies shear wave
elastography (SWE) and emerging elastomechanical hypotheses of ultrasonic
neuromodulation, both of which require accurate, efficient modeling of
radiation-force--induced tissue motion. General-purpose finite-element
elastodynamic solvers are often computationally expensive and unnecessarily
broad for shear-dominant applications. We derive a reduced viscoelastic
formulation by applying near-incompressibility, small-strain linearization,
Helmholtz decomposition, and solenoidal force projection to the full Navier
equations, yielding a Kelvin--Voigt shear wave equation that retains only the
transverse dynamics relevant to radiation-force--induced motion. An explicit
finite-difference time-domain (FDTD) implementation with second-order spatial
and temporal accuracy enforces the solenoidal body-force constraint via a
matrix-free conjugate-gradient Poisson solve. For separable radiation-force
sources, a pre-computed projection reduces this cost by one to two orders of
magnitude. In homogeneous media the solver recovers the theoretical shear
wavespeed to within $<$0.1\%, exhibits clear second-order grid convergence
(trace-$L_2$ self-convergence, $p\gtrsim2$), and matches analytical
Kelvin--Voigt attenuation and phase speed to within ${\sim}3\%$ and $<$1\% over
a 16-point viscosity sweep ($\eta\in[0,1.5]$~Pa$\cdot$s). An end-to-end RSNA
QIBA phantom benchmark recovers shear wave speeds to within ${\sim}1\%$ across a
tenfold shear-modulus range ($G=1$--$10$~kPa). The framework accommodates
spatial heterogeneity in shear modulus, density, and viscosity, and integrates
with acoustic simulators. Transcranial demonstrations through micro-CT skull
geometries produce shear displacements of 1.7--5~$\mu$m consistent with clinical
ARFI.
\end{abstract}

\section{Introduction}
\label{sec:introduction}

Ultrasound-driven mechanical phenomena in soft tissue play a central
role in several biomedical applications, most prominently shear wave
elastography~(SWE)~\cite{sarvazyan1998elasticity,barr2015elastography,gennisson2013ultrasound,sigrist2017ultrasound}
and ultrasonic
neuromodulation~\cite{blackmore2019ultrasound,jones2022neuromodulation,pinton2012braintherapy}.
In SWE, acoustic radiation
force~\cite{nightingale2002radiation,palmeri2011radiation,nyborg1965acoustic,torr1984acoustic}
generates transient elastic displacements whose propagation encodes
the underlying shear modulus of tissue. In ultrasonic neuromodulation,
related elastomechanical interactions have been proposed as potential
mechanisms linking acoustic exposure to neural
activity~\cite{blackmore2019ultrasound,kubanek2018neuromodulation,yoo2022focused}.
Both applications require accurate and computationally efficient modeling of
elastic wave generation and propagation in soft solids under ultrasound
excitation.

Extensive prior work has developed high-fidelity simulation frameworks
for SWE based on general-purpose finite element methods
(FEM)~\cite{hughes2000finite,zienkiewicz2005finite,palmeri2017fem}, including widely
used packages such as LS-DYNA and
Abaqus~\cite{bercoff2004supersonic}. These tools
can represent complex geometries, heterogeneous materials, nonlinear
constitutive behavior, and boundary interactions, making them
versatile. However, this generality comes at substantial computational
and algorithmic cost. For ultrasound-specific problems the interest lies
in small, radiation-force--induced shear displacements in nearly
incompressible soft tissue, where the full generality of large-scale FEM
is often unnecessary and can obscure the essential physics.

A central challenge in ultrasound-driven elastodynamics is the strong separation
of physical scales between acoustic and shear phenomena. Acoustic waves in soft
tissue propagate with speeds on the order of
$10^3$~m/s~\cite{szabo2014diagnostic,duck1990physical}, governed by the large
bulk modulus, whereas shear waves travel at speeds typically near
$1$--$5$~m/s~\cite{sandrin2003transient,tanter2008quantitative}. This disparity
produces temporal scales that differ by approximately three orders of magnitude.
As a result, practical modeling strategies treat the acoustic field and the
resulting radiation-force--driven elastic response on separate timescales,
coupling them through intermediate source terms. The acoustic pressure or
intensity is computed first, the radiation force is derived from it, and that
force is then supplied as a body force to a shear wave
solver~\cite{yang2018gpu}. Such sequential acousto--elastic coupling
approaches are well documented in the SWE
literature~\cite{bercoff2004supersonic,barr2015elastography}.
Full FEM modeling of ARF-induced shear wave propagation is covered by
established guidelines~\cite{palmeri2017fem}, which describe a sequential
pipeline and recommend commercial codes. That approach is general but
computationally expensive.

Numerical simulation is widely used to validate and improve shear wave imaging
techniques. FEM models have been developed for ARFI
imaging~\cite{palmeri2005fem}, for the shear wave response to radiation force
in breast~\cite{nightingale2000fem} and liver~\cite{palmeri2008liver}, for
viscoelastic propagation~\cite{lee2012viscoelastic}, and for thermal effects in
soft tissue~\cite{palmeri2004thermal}. Alternatives to full-field FEM include
the finite difference
method~\cite{orescanin2011fdtd,langdon2015}, the pseudospectral
method~\cite{qiang2015}, and the Green's function
approach~\cite{gennisson2003,bercoff2004supersonic,calle2005,chatelin2016}.
Green's function methods are well suited when the response is needed at only a
few observation points. Full-field methods (FEM, FDTD, pseudospectral) instead
compute the entire transient response over the domain.

Among finite-difference approaches, dedicated 3D FDTD simulations of shear
waves for complex modulus imaging have been
implemented~\cite{orescanin2011fdtd}. That work assumes incompressibility and
pure shear, yielding six first-order velocity--stress equations in 3D with
Kelvin--Voigt viscoelasticity on a staggered grid with PML boundaries. The
source is a vibrating needle chosen to minimize compressional energy. The
present formulation, by contrast, uses a displacement-based update with an
explicit solenoidal projection of the body force $\mathbf{f}_S$ via a scalar
Poisson (CG) solve, so that arbitrary (e.g., radiation-force) sources can be
applied without prescribing a needle geometry. Both are reduced shear-only
forward solvers. The Orescanin et~al.\ approach targets algebraic Helmholtz
inversion for complex modulus reconstruction from harmonic excitation, whereas
the present solver emphasizes integration with acoustic simulators and
Kelvin--Voigt attenuation in the shear equation.

Green's function methods offer a different trade-off. GPU-accelerated
implementations~\cite{yang2018gpu} provide fast evaluation of shear waves from
a prescribed radiation force in elastic and viscoelastic media. The present
FDTD formulation instead advances the full displacement field on a grid at each
time step, so that the response is available everywhere at once and the same
framework can accommodate spatial heterogeneity and general boundary conditions.

The present work therefore derives and implements a reduced formulation of
shear elastodynamics tailored to ultrasound-driven problems in soft solids.
Rather than pursuing full general-purpose elasticity, we isolate the minimal
physical model required to reproduce radiation-force--induced shear
displacements and their propagation while reducing computational complexity. The
formulation accommodates spatial heterogeneity in shear modulus, density, and
viscosity. We validate it against analytical benchmarks for wavespeed recovery,
grid convergence, and Kelvin--Voigt attenuation, as well as against the
standardized RSNA QIBA elastic phantom~\cite{palmeri2017fem}.
This reduction integrates tightly with acoustic simulation tools and
supports efficient multiphysics workflows relevant to SWE, elastomechanical
hypotheses of ultrasonic neuromodulation, and iterative methods for inverse
problems.

\section{Reduced Elastodynamic Formulation}
\label{sec:reduced_model}

This section derives the governing equations for ultrasound-driven shear
dynamics in soft tissue, beginning from the full linear elastodynamic continuum
description and systematically applying the physical reductions described
below. The main assumptions are near-incompressibility of soft tissue ($K \gg
\mu$), small-strain linearization, and timescale separation between acoustic
and shear dynamics.

\subsection{Near-incompressibility in soft tissue and modulus hierarchy}
\label{sec:near_incompressible}

A major simplification exploited in ultrasound-driven elastodynamics is the
extreme separation between volumetric (compressional) stiffness and shear
stiffness in soft
tissue~\cite{fung1993biomechanics,duck1990physical}. In isotropic linear
elasticity~\cite{auld1973acoustic,graff1991wave,achenbach1973wave}, the
longitudinal and shear wave speeds satisfy
\begin{align}
c_L^2 &= \frac{K+\frac{4}{3}\mu}{\rho},\\
c_S^2 &= \frac{\mu}{\rho},
\end{align}
where $\rho$ is density, $K$ is bulk modulus, and $\mu$ is shear modulus. Using
typical ultrasound values in soft
tissue~\cite{duck1990physical} ($c_L \approx 1540$~m/s and
$\rho \approx 1000$~kg/m$^3$) gives
\begin{equation}
K \approx \rho\,c_L^2 \approx 2.4\times 10^{9}\ \text{Pa},
\end{equation}
where the correction $\frac{4}{3}\mu$ is negligible because $\mu \ll K$.
Meanwhile, typical shear-wave speeds in
elastography~\cite{sandrin2003transient,barr2015elastography}
($c_S \approx 1$--$5$~m/s) imply
\begin{equation}
\mu = \rho\,c_S^2 \approx 1\times 10^{3}\text{--}2.5\times 10^{4}\ \text{Pa}.
\end{equation}
Thus, in ultrasound-relevant soft tissue,
\begin{equation}
\frac{K}{\mu} \sim 10^{5}\text{--}10^{6},
\end{equation}
which is a far stronger hierarchy than one would infer from moderate values of
Poisson's ratio alone.

For completeness, the elastic constants satisfy
\begin{equation}
E = 2\mu(1+\nu), \qquad
\nu = \frac{3K-2\mu}{2(3K+\mu)}, \qquad
K = \frac{E}{3(1-2\nu)}.
\end{equation}
Under the hierarchy $K \gg \mu$, these reduce to the familiar incompressible
limits
\begin{equation}
\nu \approx \frac{1}{2}-\frac{\mu}{6K}, \qquad
E \approx 3\mu,
\end{equation}
so that the Young's modulus remains in the kPa range even though the bulk
modulus is in the GPa range.

\paragraph{Interpretation for ultrasound-driven shear wave propagation.}
The hierarchy $K \gg \mu$ implies that compressional disturbances
propagate rapidly and are energetically expensive in volumetric
strain, whereas the radiation-force--induced response of interest in
elastography~\cite{sandrin2003transient,tanter2008quantitative} is
dominated by slower shear motion governed primarily by $\mu$. This
motivates shear-focused elastic models and sequential acousto--elastic
workflows in which the acoustic field computes the forcing while the
subsequent shear dynamics are simulated on a much slower time axis. In
ultrasound-driven problems, the external body force arises from acoustic
radiation force. Because the acoustic field evolves on timescales three
orders of magnitude faster than shear propagation, the force distribution
can be determined independently by measurement, analytical calculation, or
acoustic simulation, and then treated as a prescribed source term in
Eq.~\eqref{eq:shear_reduced_kv}.

\subsection{Full linear viscoelastodynamics}

Let $\mathbf{u}(\mathbf{x},t)$ denote the displacement field in an isotropic,
linearly viscoelastic continuum of density $\rho$. We adopt a Kelvin--Voigt
constitutive
law~\cite{catheline2004viscoelastic,ferry1980viscoelasticity}, in which the
Cauchy stress is decomposed into an elastic and a viscous contribution,
\begin{equation}
\boldsymbol{\sigma}
=
\lambda\,\mathrm{tr}(\boldsymbol{\varepsilon})\mathbf{I}
+ 2\mu\,\boldsymbol{\varepsilon}
+ 2\eta\,\dot{\boldsymbol{\varepsilon}},
\label{eq:kv_stress}
\end{equation}
where $\boldsymbol{\varepsilon} = \tfrac{1}{2}\left(\nabla \mathbf{u} + (\nabla \mathbf{u})^T\right)$
is the small-strain tensor, $\lambda$ and $\mu$ are the Lam\'e parameters, and
$\eta$ is the shear viscosity. Conservation of linear momentum with body force
density $\mathbf{f}$ yields
\begin{equation}
\rho \frac{\partial^2 \mathbf{u}}{\partial t^2}
=
\nabla\cdot \boldsymbol{\sigma} + \mathbf{f}.
\label{eq:momentum_balance}
\end{equation}
Substituting Eq.~\eqref{eq:kv_stress} into Eq.~\eqref{eq:momentum_balance}
gives
\begin{equation}
\rho \frac{\partial^2 \mathbf{u}}{\partial t^2}
=
(\lambda+\mu)\nabla(\nabla\cdot\mathbf{u})
+ \mu \nabla^2\mathbf{u}
+ \eta \nabla^2 \frac{\partial \mathbf{u}}{\partial t}
+ \mathbf{f},
\label{eq:navier_kv}
\end{equation}
where the viscous term contributes a Laplacian of particle velocity.

\subsection{Helmholtz decomposition of displacement}

The displacement field may be decomposed into irrotational (longitudinal) and
solenoidal (shear)
components~\cite{auld1973acoustic,graff1991wave},
\begin{equation}
\mathbf{u} = \nabla \phi + \nabla \times \mathbf{\Psi},
\end{equation}
where $\phi$ is a scalar potential and $\mathbf{\Psi}$ is a vector potential
satisfying $\nabla \cdot \mathbf{\Psi} = 0$.

Substituting into Eq.~\eqref{eq:navier_kv} yields decoupled wave equations for
the longitudinal and shear components,
\begin{align}
\frac{\partial^2 \phi}{\partial t^2} &= c_L^2 \nabla^2 \phi + \frac{1}{\rho} f_L,
\\
\frac{\partial^2 \mathbf{\Psi}}{\partial t^2} &= c_S^2 \nabla^2 \mathbf{\Psi} + \frac{1}{\rho} \mathbf{f}_S,
\end{align}
where $f_L$ and $\mathbf{f}_S$ denote the longitudinal and solenoidal
projections of the body force.

\subsection{Incompressible limit and reduced shear equation with viscosity}
\label{sec:incompressible_limit}

Given the extreme modulus hierarchy $K \gg \mu$
(Sec.~\ref{sec:near_incompressible}), we take the incompressible limit,
\begin{equation}
\nabla \cdot \mathbf{u} = 0,
\label{eq:incompressible}
\end{equation}
which eliminates longitudinal dynamics and introduces a pressure field $p$ as a
Lagrange multiplier that enforces the constraint. In this limit,
Eq.~\eqref{eq:navier_kv} becomes
\begin{equation}
\rho \frac{\partial^2 \mathbf{u}}{\partial t^2}
=
-\nabla p + \mu \nabla^2 \mathbf{u}
+ \eta \nabla^2 \frac{\partial \mathbf{u}}{\partial t}
+ \mathbf{f}.
\label{eq:incompressible_kv}
\end{equation}
Taking the divergence of this momentum equation and enforcing
Eq.~\eqref{eq:incompressible} yields a Poisson equation for $p$ that keeps the
displacement divergence-free at each time. Projecting
Eq.~\eqref{eq:incompressible_kv} onto the divergence-free subspace removes the
pressure gradient and gives the reduced forced shear wave equation
\begin{equation}
\rho \frac{\partial^2 \mathbf{u}}{\partial t^2}
=
\mu \nabla^2 \mathbf{u}
+ \eta \nabla^2 \frac{\partial \mathbf{u}}{\partial t}
+ \mathbf{f}_S,
\label{eq:shear_reduced_kv}
\end{equation}
where $\mathbf{f}_S$ is the solenoidal projection of the applied body force.
Equation~\eqref{eq:shear_reduced_kv} governs shear wave propagation with
wavespeed $c_S = \sqrt{\mu/\rho}$ and frequency-dependent attenuation set by
$\eta$. Under the assumptions of linear elasticity, near incompressibility,
small displacement, and timescale separation between acoustic and shear
dynamics, this equation forms the physical basis for the numerical formulation
of ultrasound-driven tissue motion.

\section{Numerical Methods}
\label{sec:numerical_methods}

\subsection{Explicit FDTD update}
Let $\bm{u}^n$ denote displacement at time $t_n=n\Delta t$. The
implemented update~\cite{taflove2005computational} is
\begin{equation}
\bm{u}^{n+1}
=
2\bm{u}^{n} - \bm{u}^{n-1}
+
\Delta t^2 \, \bm{a}^{n},
\end{equation}
with acceleration
\begin{equation}
\bm{a}^{n}
=
\frac{\mu}{\rho}\nabla_h^2 \bm{u}^{n}
+
\frac{\eta}{\rho}\nabla_h^2 \dot{\bm{u}}^{n}
+
\frac{1}{\rho}\bm{f}_S^{n},
\quad
\dot{\bm{u}}^{n}\approx\frac{\bm{u}^{n}-\bm{u}^{n-1}}{\Delta t}.
\end{equation}
$\nabla_h^2$ is a second-order finite-difference 3D Laplacian (7-point stencil)
on cell centers.

For homogeneous media without viscosity, a standard stability bound is
\begin{equation}
\Delta t \lesssim \frac{1}{c_s}
\left(
\frac{1}{\Delta x^2}+\frac{1}{\Delta y^2}+\frac{1}{\Delta z^2}
\right)^{-1/2},
\end{equation}
which motivates the CFL-based~\cite{courant1928partial} step choice in
validation scripts. The overall scheme is second-order accurate in time for the
elastic component and first-order accurate in time for viscous dissipation,
consistent with Kelvin--Voigt
damping~\cite{catheline2004viscoelastic}. Numerical convergence and stability
are verified empirically in Sec.~\ref{sec:results} through grid-refinement
and timestep studies.

\subsection{Helmholtz projection via conjugate gradient}
\label{sec:helmholtz_projection}

The reduced shear equation~\eqref{eq:shear_reduced_kv} requires the solenoidal
projection $\bm{f}_S$ of the applied body force. In the general solver, a
different body force can be supplied at each time step, and the solenoidal
projection is computed independently for each one. This captures fully
time-dependent body forces, such as those arising from a moving acoustic focus
or time-varying pulses.

\subsubsection{Solenoidal projection of body force.}
At each time step $t_n$, the body force $\bm{f}^n$ is decomposed into
solenoidal and irrotational parts,
\begin{equation}
\bm{f}^n = \bm{f}_S^{\,n} + \nabla \phi^n,
\quad
\nabla\cdot\bm{f}_S^{\,n}=0.
\end{equation}
Taking divergence gives a scalar Poisson equation
\begin{equation}
\nabla^2 \phi^n = \nabla\cdot\bm{f}^n,
\end{equation}
from which the solenoidal component is recovered as
\begin{equation}
\bm{f}_S^{\,n} = \bm{f}^n - \nabla\phi^n.
\end{equation}
This projection removes forcing components that would excite compressional
motion. The solver is therefore a shear-mode propagator, not a full
elastic-wave solver.

\subsubsection{Choice of conjugate gradient for the scalar Poisson problem.}
Only one scalar field $\phi$ must be solved at each step (rather than a full
vector system). After finite-difference discretization, this becomes a large
sparse linear system $A\phi=b$, where $A$ is the discrete negative Laplacian
on interior nodes. Conjugate gradient (CG)~\cite{shewchuk1994conjugate,golub2013matrix} is a
natural choice because $A$ is symmetric positive-definite on the interior
unknowns, matching the theoretical requirements of CG. The 3D problem is large
and sparse, making direct factorization memory-prohibitive, whereas CG requires
only repeated stencil-like matrix--vector operations that scale well in three
dimensions.

\subsubsection{Numerical implementation details.}
The projection solve is performed in matrix-free form as
\begin{equation}
A\phi=b, \qquad A=-\nabla_h^2, \qquad b=\nabla_h\cdot\bm{f},
\end{equation}
where $\nabla_h^2$ is a second-order central-difference 3D Laplacian (7-point
stencil on interior nodes).

The code uses an interior mask $m$ (1 in interior, 0 on outer layer) and
applies it after each update,
\begin{equation}
\phi \leftarrow \phi \odot m,\quad r \leftarrow r \odot m,\quad p \leftarrow p \odot m.
\end{equation}
This enforces the chosen boundary treatment and avoids drift of the boundary
unknowns.

Starting from $\phi_0=0$,
\begin{align}
r_0 &= (b-A\phi_0)\odot m, \qquad p_0=r_0,\\
\alpha_k &= \frac{r_k^T r_k}{p_k^T A p_k},\\
\phi_{k+1} &= (\phi_k + \alpha_k p_k)\odot m,\\
r_{k+1} &= (r_k - \alpha_k A p_k)\odot m,\\
\beta_k &= \frac{r_{k+1}^T r_{k+1}}{r_k^T r_k},\\
p_{k+1} &= (r_{k+1} + \beta_k p_k)\odot m.
\end{align}
No sparse matrix is assembled explicitly. $Aq$ is computed by applying the
Laplacian stencil to $q$.

The solver terminates when the relative residual drops below tolerance,
\begin{equation}
\sqrt{\frac{\|r_k\|_2^2}{\|r_0\|_2^2}} < \text{tol},
\end{equation}
or when the maximum number of iterations is reached. Additional guards stop the
iteration if the denominator or residual becomes non-finite or non-positive.

\subsection{Pre-computed projection for separable body forces}
\label{sec:precomputed_projection}

In the common case of acoustic radiation force (ARF) driven shear wave
excitation, the spatial and temporal dependences of the body force are
separable. The ARF spatial distribution $\bm{b}_0(\bm{x})$ is determined
entirely by the acoustic intensity field and tissue absorption, while the
temporal dependence is controlled by the push envelope. The body force then
takes the form
\begin{equation}
\bm{f}(\bm{x},t) = \bm{b}_0(\bm{x})\,g(t),
\label{eq:separable_force}
\end{equation}
where $g(t)$ is a scalar temporal envelope (for example, a step function that
equals one during the push and zero afterwards). Because the Helmholtz--Hodge
decomposition is linear, the solenoidal projection commutes with the scalar
envelope,
\begin{equation}
\bm{f}_S^{\,n}
= \mathcal{P}_S\!\bigl[\bm{b}_0\,g(t_n)\bigr]
= g(t_n)\,\mathcal{P}_S[\bm{b}_0]
= g(t_n)\,\bm{f}_{S,0},
\label{eq:projected_separable}
\end{equation}
where $\bm{f}_{S,0} \equiv \mathcal{P}_S[\bm{b}_0]$ is the divergence-free
projection of the spatial pattern.

\subsubsection{Algorithm.}
The pre-computed projection proceeds in two stages.

\begin{enumerate}[leftmargin=1.5em]
\item \textbf{Pre-computation.} For a given spatial body force pattern
$\bm{b}_0$, solve the scalar Poisson problem
\begin{equation}
\nabla_h^2 \phi_0 = \nabla_h \cdot \bm{b}_0,
\end{equation}
via conjugate gradient, then form the solenoidal projection
\begin{equation}
\bm{f}_{S,0} = \bm{b}_0 - \nabla_h \phi_0.
\end{equation}
This step is repeated each time the spatial force distribution changes.

\item \textbf{Time stepping.} While the spatial pattern remains fixed, the
solenoidal force at each step $n$ follows by scalar multiplication,
\begin{equation}
\bm{f}_S^{\,n} = g(t_n)\,\bm{f}_{S,0}.
\end{equation}
The FDTD update then proceeds with the acceleration
\begin{equation}
\bm{a}^{n}
=
\frac{\mu}{\rho}\nabla_h^2 \bm{u}^{n}
+
\frac{\eta}{\rho}\nabla_h^2 \dot{\bm{u}}^{n}
+
\frac{g(t_n)}{\rho}\,\bm{f}_{S,0}.
\end{equation}
\end{enumerate}

In the standard ARF scenario the acoustic intensity field is computed once and
the resulting spatial force pattern $\bm{b}_0(\bm{x})$ is fixed for the entire
simulation, so the pre-computation is performed only once. More generally, the
projection can be updated whenever the spatial pattern changes (e.g., between
successive pushes or insonications at different foci), with only $O(N)$ scaling
required during the intervening time steps.

Each pre-computation requires one CG solve at cost
$O(N \cdot n_{\mathrm{iter}})$, while the per-step scaling costs only $O(N)$.
When the spatial pattern is fixed, the total projection cost is
$O(N \cdot n_{\mathrm{iter}} + N \cdot N_t)$ compared with
$O(N \cdot n_{\mathrm{iter}} \cdot N_t)$ for the general per-step solver.
For practical problem sizes, for example $N = 106^3 \approx 1.2 \times 10^6$
points and $N_t = 85$ steps, the pre-computed projection reduces total
simulation time from $O(10^2\text{--}10^3)$~s to $O(10)$~s, an improvement of
one to two orders of magnitude. Both the general per-step solver and the
pre-computed variant are implemented. The choice is made automatically based on
whether the user supplies a separable source.

\subsection{Sequential acoustic--elastic coupling}
\label{sec:coupling}

Because the acoustic field evolves on timescales three orders of magnitude
faster than shear propagation, the two problems can be solved sequentially. The
acoustic pressure and particle velocity fields are computed first, a body force
distribution is derived from them, and this force is supplied as a prescribed
source term to Eq.~\eqref{eq:shear_reduced_kv}. This sequential coupling is the
standard workflow in SWE simulation~\cite{palmeri2017fem,yang2018gpu}.

The acoustic field can be obtained by measurement, by analytical calculation, or
from any suitable numerical solver such as Field II, Fullwave~\cite{pinton2020fullwave},
or the angular spectrum approach~\cite{zeng2008evaluation}.

Two formulations compute the acoustic radiation body force from the acoustic
field. The plane-wave approximation gives a scalar force along the beam
axis~\cite{nyborg1965acoustic,torr1984acoustic},
\begin{equation}
  b_0 = \frac{2\,\alpha_{\mathrm{Np}}}{c}\,I_{\mathrm{SPPA}},
  \qquad
  I_{\mathrm{SPPA}} = \frac{1}{\rho\, c}
    \;\frac{\sum_{i=1}^{N_{\mathrm{total}}} p_i^2}{N_{\mathrm{pulse}}},
  \label{eq:isppa_norm}
\end{equation}
where $\alpha_{\mathrm{Np}}$ is the absorption in Np/m at the transmit
frequency. The plane-wave approximation is easier to implement for non-vectorial
estimates of the pressure wavefield. However, for converging beams the
plane-wave assumption breaks down near the focal zone, and a more accurate
approach uses the time-averaged Poynting vector $\bm{I} = \langle p\, \bm{v} \rangle$,
\begin{equation}
  f_i = \frac{2\,\alpha_{\mathrm{Np}}}{c}\,I_i,
  \qquad
  I_i = \langle p\, v_i \rangle,
  \qquad i \in \{x,y,z\},
  \label{eq:arf_poynting}
\end{equation}
yielding three spatially varying force components.


\subsection{Boundary and initial conditions}
\label{sec:bc_ic}

In the current implementation, the computational domain is terminated with a
fixed (Dirichlet) boundary condition by setting displacement to zero on the
outer layer. This choice simplifies the staggered-grid update and provides a
well-posed finite-domain problem, but it can introduce reflections that
contaminate late-time arrivals. Absorbing boundary treatments such as perfectly
matched layers
(PML)~\cite{berenger1994perfect,collino2001application} can reduce such
artifacts when late-time fidelity is required. Accordingly, validation and
analysis are restricted to early and intermediate propagation times, or to
sensor offsets not dominated by boundary interactions. The initial
state is commonly specified as a divergence-free velocity field constructed from
a vector potential (a curl-based initialization), which preferentially excites
shear motion while suppressing compressional content.

\section{Results}
\label{sec:results}

\subsection{Non-viscous wavespeed recovery}
\label{sec:nonvisc_validation}

To isolate elastic wavespeed recovery, the first validation disables viscosity
($\eta=0$) and uses a curl-based vector-potential source to excite a
divergence-free initial velocity field. This source is constructed from a
Gaussian envelope in the $x$--$y$ plane, applied through the curl of a vector
potential so that the resulting velocity field is solenoidal by construction.
The curl initialization preferentially excites shear motion while suppressing
compressional content, providing a test of the shear propagation numerics
without confounds from longitudinal waves or from the Helmholtz projection
step.

\begin{figure}
\centering
\includegraphics[width=0.95\textwidth]{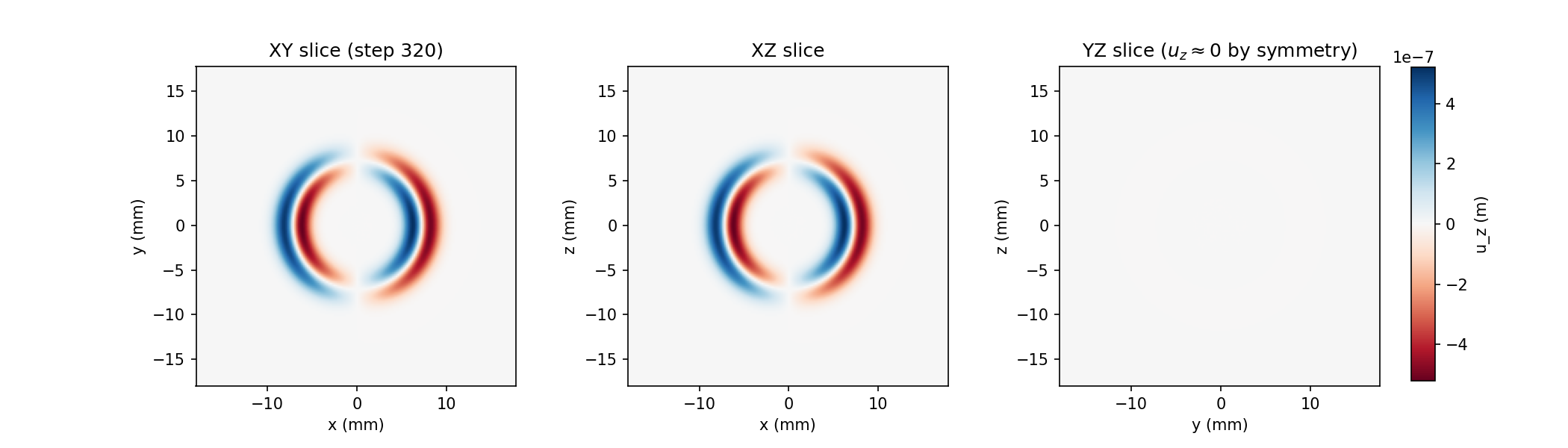}\\[6pt]
\includegraphics[width=0.45\textwidth]{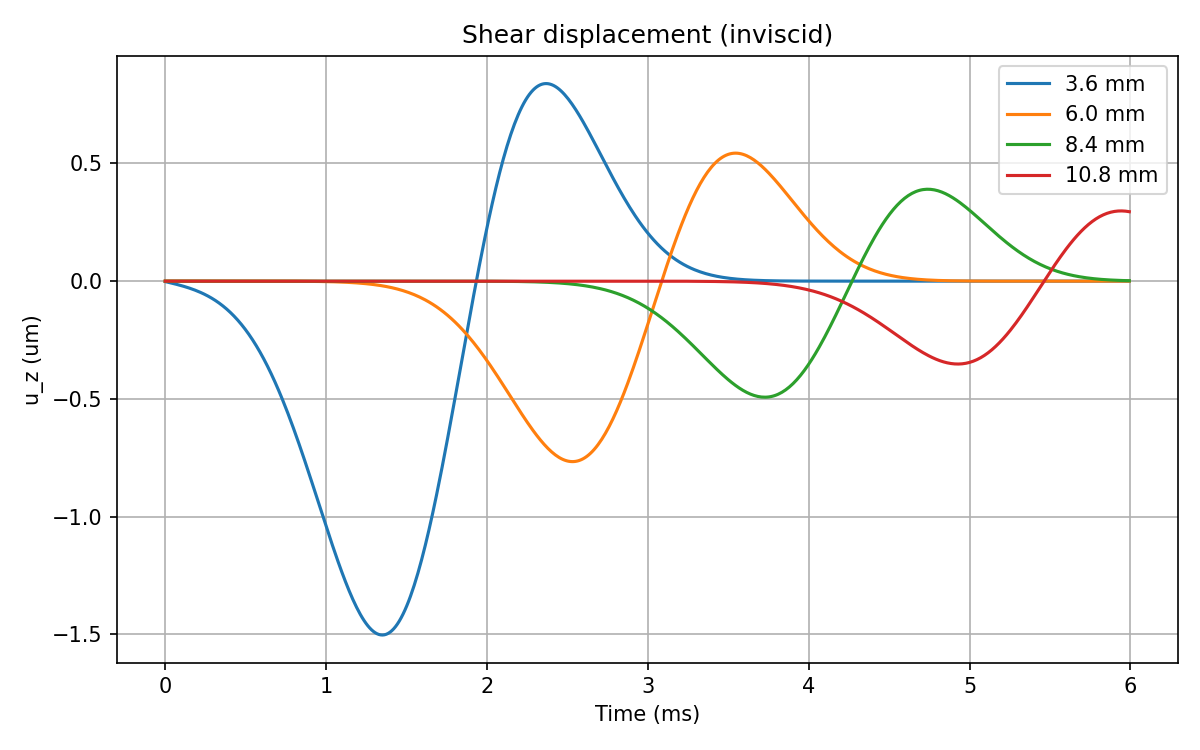}
\caption{Non-viscous wavespeed validation. \textbf{Top:} Shear-wave field
snapshot ($u_z$), $x$--$y$ slice (left), $x$--$z$ slice (middle), $y$--$z$
slice at $x=0$ (right). \textbf{Bottom:} Displacement traces ($u_z$) at four
sensor offsets (3.6, 6.0, 8.4, 10.8~mm). Wavespeed is estimated from peak
arrival times.}
\label{fig:nonvisc_validation}
\end{figure}

The simulation domain was implemented as a cubic $180^3$ grid ($L_x=L_y=L_z=36$~mm,
$\Delta x=0.2$~mm) with homogeneous soft-tissue parameters
($\rho=1000$~kg/m$^3$, $c_s=2.0$~m/s). Figure~\ref{fig:nonvisc_validation} shows the resulting shear-wave
displacement field ($u_z$) at a representative time step in three orthogonal
slice planes, along with $u_z$ traces recorded at four sensors placed at
offsets of 3.6, 6.0, 8.4, and 10.8~mm from the source along the depth axis.
The wave propagates outward in the $x$--$y$ and $x$--$z$ planes as expected.
The $y$--$z$ slice at $x=0$ is near zero because the curl-based initialization
drives motion in the $x$--$z$ plane, and the center $x$-plane is a nodal surface
for $u_z$ by symmetry. Wavespeed was estimated from peak arrival times in the
sensor traces.

The target shear speed is 2.000~m/s (from $\mu=\rho c_s^2$). The mean
recovered speed is 1.976~m/s (${\sim}$1.2\% below
target) and the distance--time linear-fit speed is 1.998~m/s ($<$0.1\% error),
confirming that the shear-wave propagation physics and numerics are correct in
this non-viscous benchmark regime.

\subsection{Grid convergence study}
\label{sec:convergence}

An inviscid convergence sweep was performed with fixed physical domain size
($L_x=L_y=L_z=36$~mm) and fixed physical sensor distances, over a five-level
refinement range $n\in\{60,90,120,150,180\}$ chosen so that the sensor
distances fall exactly on grid nodes at every level. So that each grid
discretizes the \emph{same} continuum problem, the Gaussian excitation is held
at a fixed \emph{physical} width ($\sigma_{xy}=3$~mm, i.e.\ $5$ grid points at
the coarsest level and proportionally more as the grid refines) and the
velocity-damping coefficient is scaled per time step so that the cumulative
damping over the fixed time window is grid-independent. Holding the source width
at a fixed number of grid points instead, so that its physical size shrinks
under refinement, conflates discretization error with a changing source, and is
avoided here.

\begin{figure}
\centering
\includegraphics[width=0.6\textwidth]{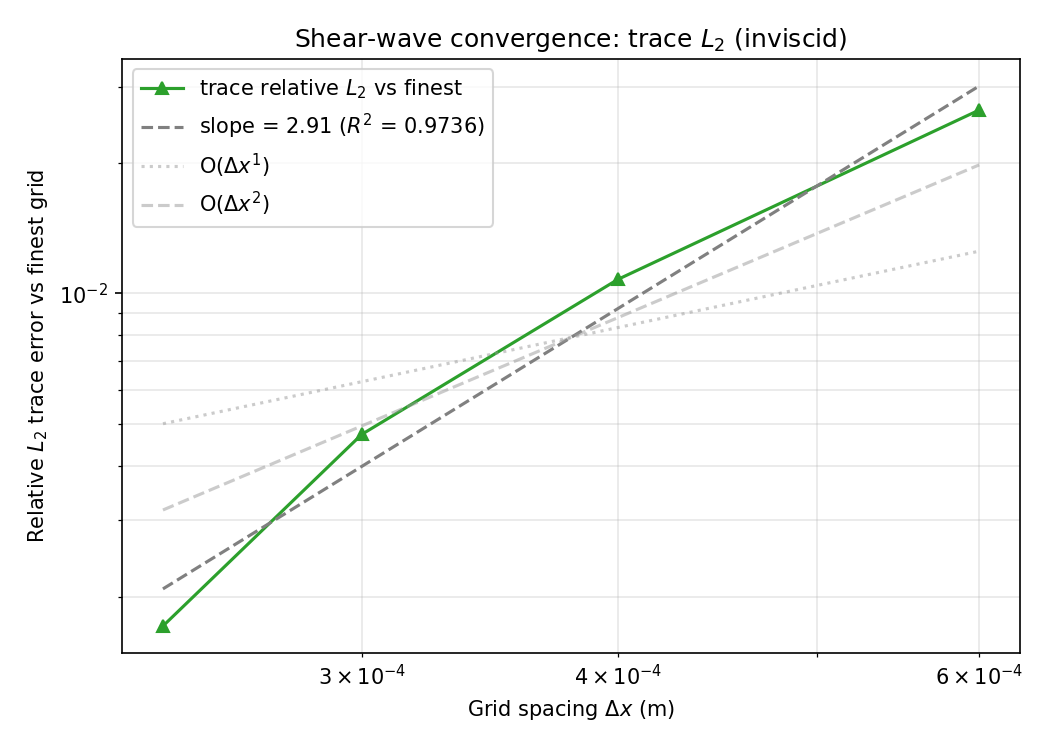}
\caption{Grid convergence of the full-waveform trace $L_2$ self-convergence
error (relative $L_2$ error of velocity traces vs the finest grid) against grid
spacing (log--log), for a fixed $3$~mm physical source. The fitted global slope
is $p\approx2.9$ ($R^2=0.97$). Dashed reference lines show $O(\Delta x)$ and
$O(\Delta x^2)$.}
\label{fig:convergence}
\end{figure}

Convergence is quantified by the full-waveform \emph{trace $L_2$ error}, the
relative $L_2$ norm of the velocity-trace difference at each sensor
(interpolated onto a common time base), averaged across sensors, measured
against the finest grid ($n=180$) as a self-convergence reference. Unlike a
scalar arrival-time speed, it captures the whole waveform.
Table~\ref{tab:convergence} lists the error at each level and
Fig.~\ref{fig:convergence} shows it against grid spacing. The error falls
monotonically from $2.7\%$ to $0.2\%$. Fitting
trace-$L_2 \propto \Delta x^p$ in log--log space gives a global slope
$p=2.9$ ($R^2=0.97$). The well-separated levels give per-interval orders
$p\approx2.2$--$2.9$, with the pair nearest the reference inflated by the
self-convergence reference (a known bias of self-convergence studies). This
confirms convergence at least as fast as the formal second-order accuracy of the
7-point Laplacian stencil and leapfrog time integration.

\begin{table}
\caption{Grid convergence of the full-waveform trace $L_2$ self-convergence
  error (relative $L_2$ norm of the velocity traces against the finest grid,
  $n=180$, fixed $36$~mm domain, fixed $3$~mm physical source).}
\centering
\begin{tabular}{c c c}
\hline
$n$ & $\Delta x$ ($\mu$m) & trace $L_2$ error \\
\hline
60  & 600 & 2.65\% \\
90  & 400 & 1.08\% \\
120 & 300 & 0.47\% \\
150 & 240 & 0.17\% \\
180 & 200 & --- (reference) \\
\hline
\end{tabular}
\label{tab:convergence}
\end{table}

We deliberately do \emph{not} use the arrival-time wavespeed error as a
grid-convergence metric. With the source width held physically fixed, the
peak-picked wavespeed reflects a source-width- and near-field-dependent
estimation bias that does not vanish under refinement, so it is not a clean
measure of discretization error. The accuracy of the recovered wavespeed itself
is instead established by the non-viscous benchmark
(Fig.~\ref{fig:nonvisc_validation}, linear-fit speed within $<$0.1\%).

\subsection{Kelvin--Voigt attenuation benchmark}
\label{sec:kv_validation}

To avoid known bias from finite 3D pulse-domain effects, viscosity validation
uses a dedicated 1D harmonic benchmark, which better matches Kelvin--Voigt
analytic assumptions (single frequency, effectively plane-wave propagation,
homogeneous medium). To align with Kelvin--Voigt theory and cover a soft tissue range, the
benchmark was run on $\eta\in[0,1.5]$~Pa$\cdot$s with
$\Delta\eta=0.1$~Pa$\cdot$s. The high-$\eta$ configuration uses adaptive
sensor placement based on the expected e-folding length, adaptive drive-amplitude
scaling, and a smooth frequency schedule $f_0(\eta)$ guided by
dimensional analysis.

Define the loss parameter and dimensionless attenuation
\begin{equation}
\mathrm{De}_v=\frac{\omega\eta}{\mu}, \qquad
\Pi_\alpha=\frac{\alpha c_s}{\omega}, \qquad c_s=\sqrt{\mu/\rho}.
\end{equation}
For weak-to-moderate loss ($\mathrm{De}_v\ll 1$),
\begin{equation}
\Pi_\alpha \approx \frac{\mathrm{De}_v}{2}
\;\Rightarrow\;
\alpha \approx \frac{\eta \omega^2}{2\rho c_s^3}.
\label{eq:alpha_scaling}
\end{equation}

A key challenge for attenuation measurement at high viscosity is that the
e-folding length $\ell_e=1/\alpha$ shrinks rapidly. At $\eta=1.5$~Pa$\cdot$s
the theoretical $\alpha\approx 824$~Np/m gives $\ell_e\approx 1.2$~mm. Sensors
placed beyond a few e-folding lengths measure only noise, biasing the
attenuation fit low. Sensor positions are thus adapted per $\eta$ based on the
theoretical attenuation at the scheduled drive frequency.

\begin{figure}
\centering
\includegraphics[width=0.49\textwidth]{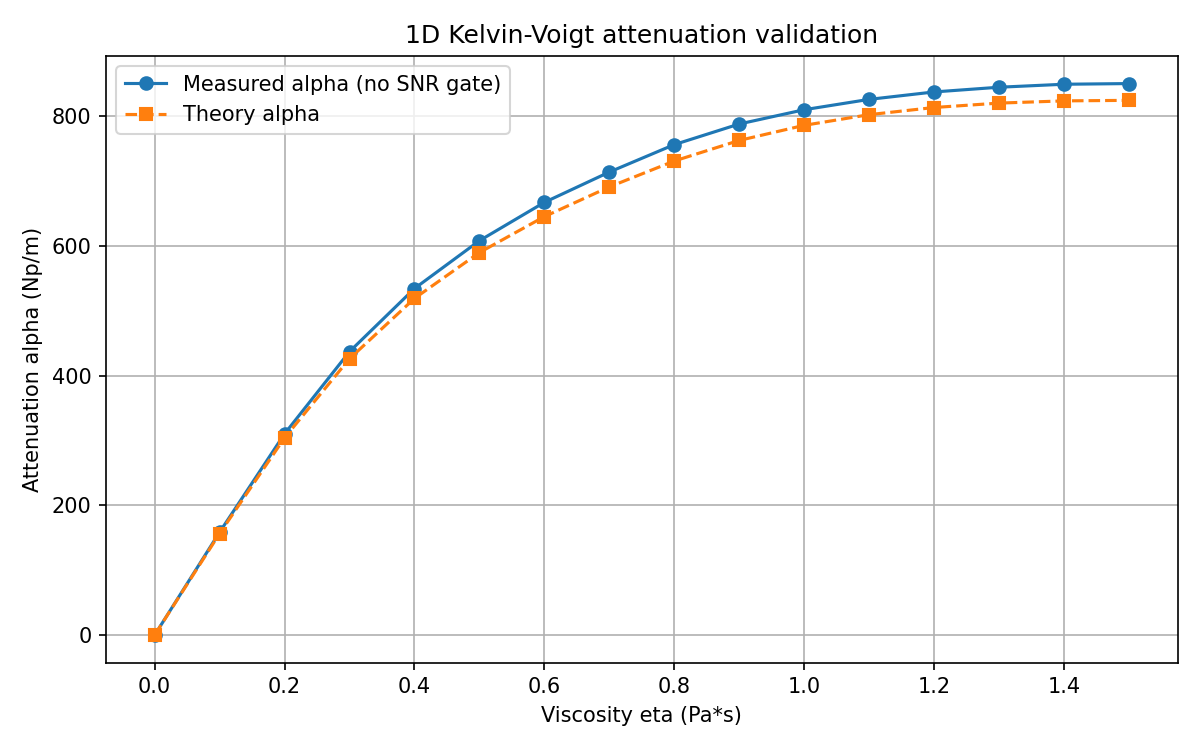}%
\hfill
\includegraphics[width=0.49\textwidth]{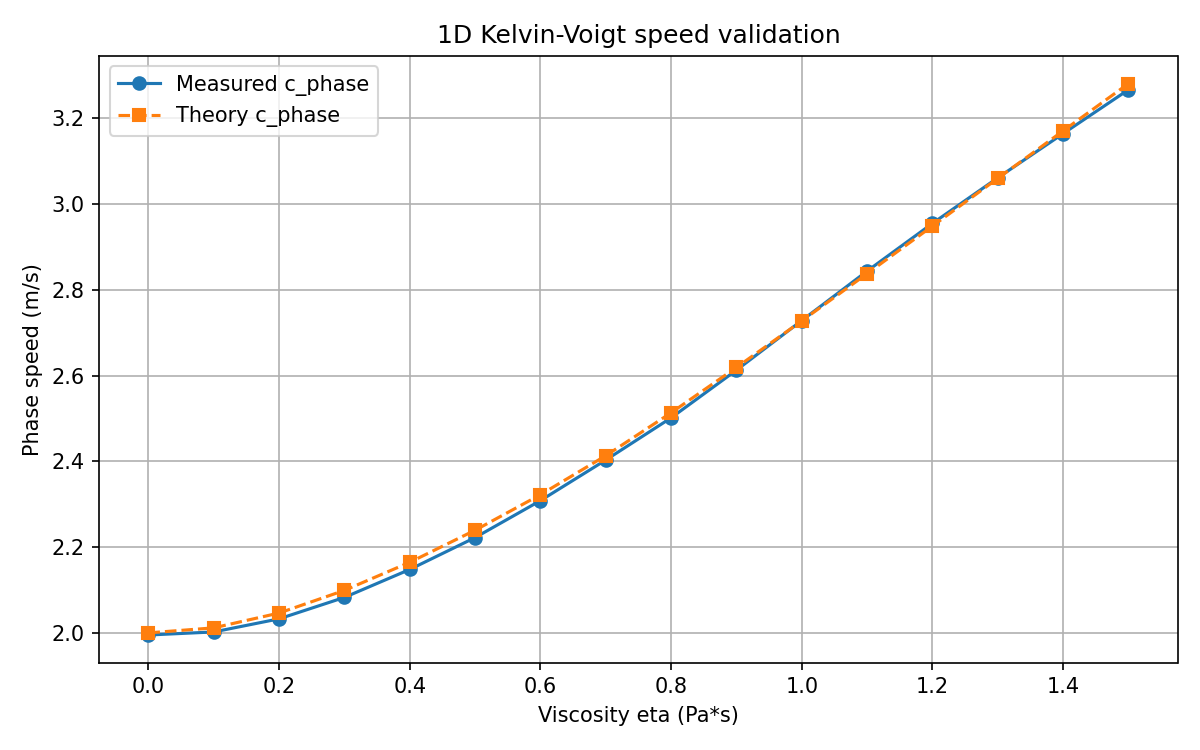}
\caption{1D harmonic Kelvin--Voigt benchmark over
$\eta\in[0,1.5]$~Pa$\cdot$s with adaptive sensor placement and frequency
scheduling. \textbf{Left:} Attenuation (measured vs theory).
\textbf{Right:} Phase speed (measured vs theory).}
\label{fig:kv1d}
\end{figure}

Table~\ref{tab:kv_results} presents representative results from the 16-point
viscosity sweep.  Across the viscosity sweep (attenuation quoted for $\eta\ge0.1$~Pa$\cdot$s,
where it is resolvable), the
measured attenuation tracks the Kelvin--Voigt prediction with a mean absolute
error of ${\sim}$3\%, while the phase speed agrees to within $<$1\%.  Both
measured quantities increase monotonically with viscosity, consistent with the
analytical dispersion relation. The results confirm accurate recovery of the
analytical Kelvin--Voigt attenuation and dispersion across the full viscosity
range.

\begin{table}
\caption{1D harmonic Kelvin--Voigt benchmark, measured vs.\ theoretical
  attenuation and phase speed at selected viscosities.  Errors are relative
  to the exact Kelvin--Voigt dispersion relation at the scheduled drive
  frequency.}
\centering
\begin{tabular}{c c c c c c c}
\hline
$\eta$ & $\alpha_{\mathrm{meas}}$ & $\alpha_{\mathrm{th}}$ & $\alpha$ error
       & $c_{\mathrm{meas}}$ & $c_{\mathrm{th}}$ & $c$ error \\
(Pa$\cdot$s) & (Np/m) & (Np/m) & (\%) & (m/s) & (m/s) & (\%) \\
\hline
0.10 & 158.5 & 156.4 & $+1.3$ & 2.002 & 2.012 & $-0.5$ \\
0.30 & 436.8 & 425.8 & $+2.6$ & 2.083 & 2.099 & $-0.8$ \\
0.50 & 607.8 & 589.5 & $+3.1$ & 2.222 & 2.239 & $-0.8$ \\
1.00 & 809.7 & 785.6 & $+3.1$ & 2.729 & 2.728 & $<0.1$ \\
1.50 & 850.0 & 824.2 & $+3.1$ & 3.267 & 3.281 & $-0.4$ \\
\hline
\end{tabular}
\label{tab:kv_results}
\end{table}

\subsection{QIBA elastic phantom benchmark}
\label{sec:qiba_validation}

The preceding validations use idealized source configurations (curl-based
impulse, harmonic plane wave) to isolate individual solver properties.  This
section validates the complete end-to-end pipeline (acoustic simulation,
radiation force computation, and shear wave propagation) against the
standardized digital phantoms defined by the RSNA QIBA Ultrasound Shear Wave
Speed Biomarker Committee~\cite{palmeri2017fem}.  These phantoms were
designed for inter-solver comparison and have been validated with LS-DYNA
and Abaqus finite element codes.

Four homogeneous elastic phantoms are simulated with shear moduli
$G\in\{1, 2, 5, 10\}$~kPa, density $\rho=1000$~kg/m$^3$, and Poisson's ratio
$\nu=0.495$ (nearly incompressible), matching the QIBA specification.  The
acoustic radiation force is modeled as a three-dimensional Gaussian body force
following the convention used in the QIBA FEM
pipeline~\cite{palmeri2017fem},
\begin{equation}
  b(\bm{x}) = A\exp\!\Bigl[-\Bigl(\frac{x-x_f}{\sigma_x}\Bigr)^{\!2}
  - \Bigl(\frac{y}{\sigma_y}\Bigr)^{\!2}
  - \Bigl(\frac{z}{\sigma_z}\Bigr)^{\!2}\Bigr],
  \label{eq:gaussexc}
\end{equation}
where $x_f=30$~mm is the focal depth and the $1/e$ widths are derived from the
beam physics of the QIBA curvilinear array ($f/2.0$, 3~MHz), namely
$\sigma_y \approx 0.62$~mm (lateral),
$\sigma_x \approx 4.3$~mm (axial),
and $\sigma_z \approx 1.1$~mm (elevational, from the 50~mm lens focus).
The push duration is 167~$\mu$s (500 cycles at 3~MHz).  The computational grid
uses $\Delta x = 0.25$~mm with a domain of $60\times50\times50$~mm$^3$
($240\times200\times200$ nodes).  For each phantom, the solver is run with
$\eta=0$ (pure elastic) and the shear wave group velocity is estimated by
cross-correlation of velocity traces at lateral offsets of 4--14~mm from the
push axis.

\begin{figure}
\centering
  \includegraphics[width=\textwidth]{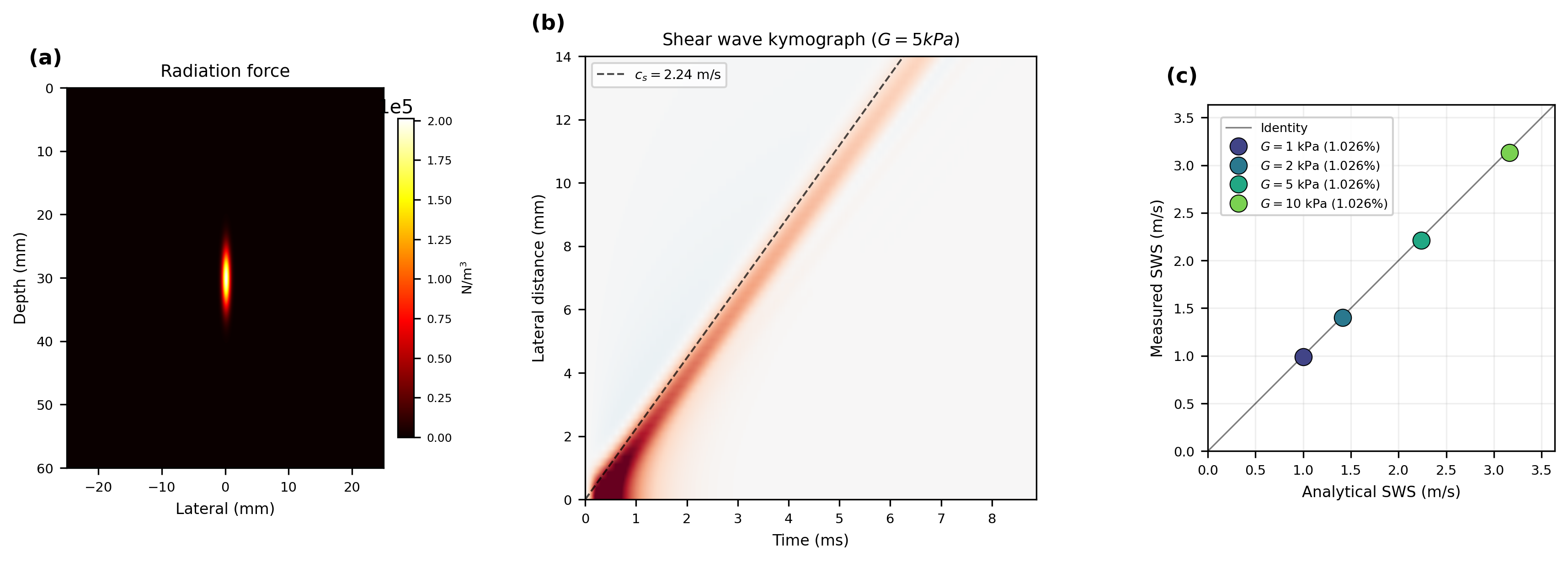}
  \caption{QIBA elastic phantom validation using a Gaussian body force
  excitation matching the QIBA convention.
  \textbf{(a)}~Gaussian radiation force distribution
  (Eq.~\ref{eq:gaussexc}), with the focal spot centered at 30~mm depth.
  \textbf{(b)}~Space--time kymograph of axial displacement for the
  $G=5$~kPa phantom, where the dashed line indicates the analytical shear wave
  speed $c_s=2.24$~m/s. \textbf{(c)}~Recovered shear wave speed vs.\
  analytical value for all four QIBA elastic phantoms ($G=1$--$10$~kPa),
  all configurations recover the wavespeed to within ${\sim}$1\%.}
  \label{fig:qiba_validation}
\end{figure}

Figure~\ref{fig:qiba_validation}(a) shows the computed radiation force
distribution, exhibiting the characteristic tight focal spot of an $f/2.0$
focused beam.  The space--time kymograph in
Figure~\ref{fig:qiba_validation}(b) displays the displacement field for the
$G=5$~kPa phantom as a function of lateral distance and time. The shear wave
front propagates at a slope consistent with the analytical speed
$c_s=\sqrt{G/\rho}=2.24$~m/s (dashed line).
Figure~\ref{fig:qiba_validation}(c) summarizes the recovered shear wave speed
for all four phantoms.  Errors relative to the analytical value are
${\sim}1.0\%$ for all four configurations ($G=1$--$10$~kPa), confirming
accurate wavespeed recovery across a tenfold range of shear moduli.  This
level of agreement is well within the inter-solver variability reported in the
QIBA round-robin comparison~\cite{palmeri2017fem}.

\section{Transcranial Application Demonstrations}
\label{sec:transcranial_demos}

Having validated the shear wave solver on homogeneous benchmark media, this
section applies a complete acousto--elastic pipeline
(Section~\ref{sec:coupling}) to transcranial ultrasound through
a realistic human skull. Two transducer configurations are considered, namely a
1024-element sparse imaging array emitting an ARF push~\cite{mccall2023sparse} and the focused ultrasound Philips TIPS curved-bowl transducer emitting a neuromodulatory sequence.
Both simulations use the same \emph{ex~vivo} micro-CT skull specimen (Halle
temporal bone)~\cite{kirchner2022skull} and share the same four-stage workflow,
comprising acoustic transmit through the skull (GPU FDTD), computation of the
acoustic radiation force, shear wave FDTD simulation of viscoelastic
brain-tissue displacement, and visualization of all intermediate and final
fields.

\subsection{Sparse phased array through micro-CT skull}
\label{sec:sparse_demo}

\begin{figure}
  \centering
  \includegraphics[width=\textwidth]{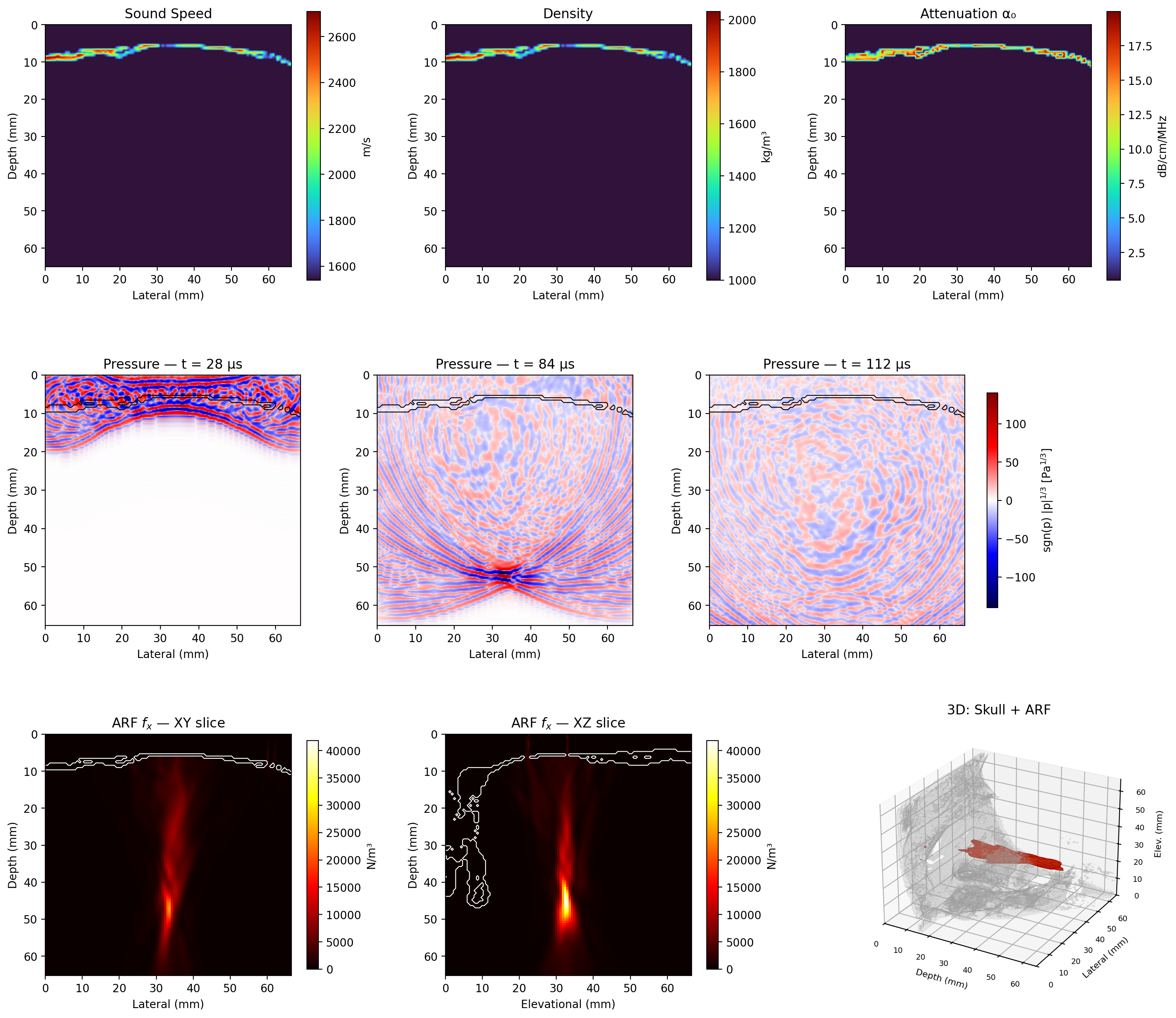}
  \caption{Acoustical medium, pressure propagation, and acoustic radiation
    force for the sparse array transcranial configuration.
    \textbf{Top row:} Sound speed, density, and attenuation coefficient
    $\alpha_0$ derived from micro-CT skull data ($x$--$y$ slices at full
    resolution).
    \textbf{Middle row:} Pressure propagation snapshots showing the focused
    beam emerging from the transducer, propagating through the skull (black
    contour), and converging at the focal zone.
    \textbf{Bottom row:} Depth-component ARF ($f_x$) in $x$--$y$ and $x$--$z$
    slices (brain-masked, skull contour in white), with a 3D rendering of the
    ARF isosurface (red) within the skull geometry (gray).}
  \label{fig:acoustical_and_pressure}
\end{figure}

The first demonstration uses a 1024-element sparse phased array with
750~$\mu$m element pitch~\cite{mccall2023sparse}. Element positions are loaded from an HDF5 connector
file and mapped onto the simulation grid ($66 \times 66$~mm
lateral--elevational aperture). Per-element focusing delays steer the beam to a
50~mm depth focal point.
The heterogeneous skull medium is constructed from a micro-CT NRRD volume of
the Halle temporal bone specimen~\cite{kirchner2022skull}. Hounsfield units are mapped to spatially
varying sound speed (1540--2900~m/s), density (1000--2200~kg/m$^3$), and
attenuation coefficient following established CT-to-acoustic
mappings~\cite{connor2002skull,pinton2012braintherapy}. The background tissue
is assigned standard soft-tissue properties ($c = 1540$~m/s,
$\rho = 1000$~kg/m$^3$,
$\alpha_0 = 0.5$~dB\,cm$^{-1}$\,MHz$^{-1}$)~\cite{duck1990physical}. The
acoustic simulation uses the Fullwave GPU FDTD
solver~\cite{pinton2009fullwave} with
exponential attenuation at $f_0 = 1$~MHz, source pressure amplitude of
1.5~MPa, and a sparse-grid sensor (4$\times$ downsampling in each spatial
dimension). Figure~\ref{fig:acoustical_and_pressure} shows the acoustical
medium property maps and three pressure propagation snapshots as the focused
beam propagates through the skull and converges at the 50~mm focal zone.
The velocity-based Poynting vector formulation
Eq.~\eqref{eq:arf_poynting} is used with pulse normalization corrections to
compute the acoustic radiation force.
Non-brain regions (coupling medium, skull interior, and
pre-skull tissue) are masked to zero so that only the intracranial force field
drives the shear wave simulation.
The depth component gives $f_x^{\max} \approx 41{,}800$~N/m$^3$, with
lateral and elevational components $|f_y|^{\max} \approx 7{,}000$~N/m$^3$
and $|f_z|^{\max} \approx 6{,}400$~N/m$^3$.
Figure~\ref{fig:acoustical_and_pressure} (bottom row) shows the depth component
of the resulting body force distribution. The pre-skull coupling medium and
skull interior are masked so that only the brain-side force field is displayed.
A 3D rendering of the ARF isosurface within the skull geometry is included for
spatial context.

\begin{figure}
  \centering
  \includegraphics[width=\textwidth]{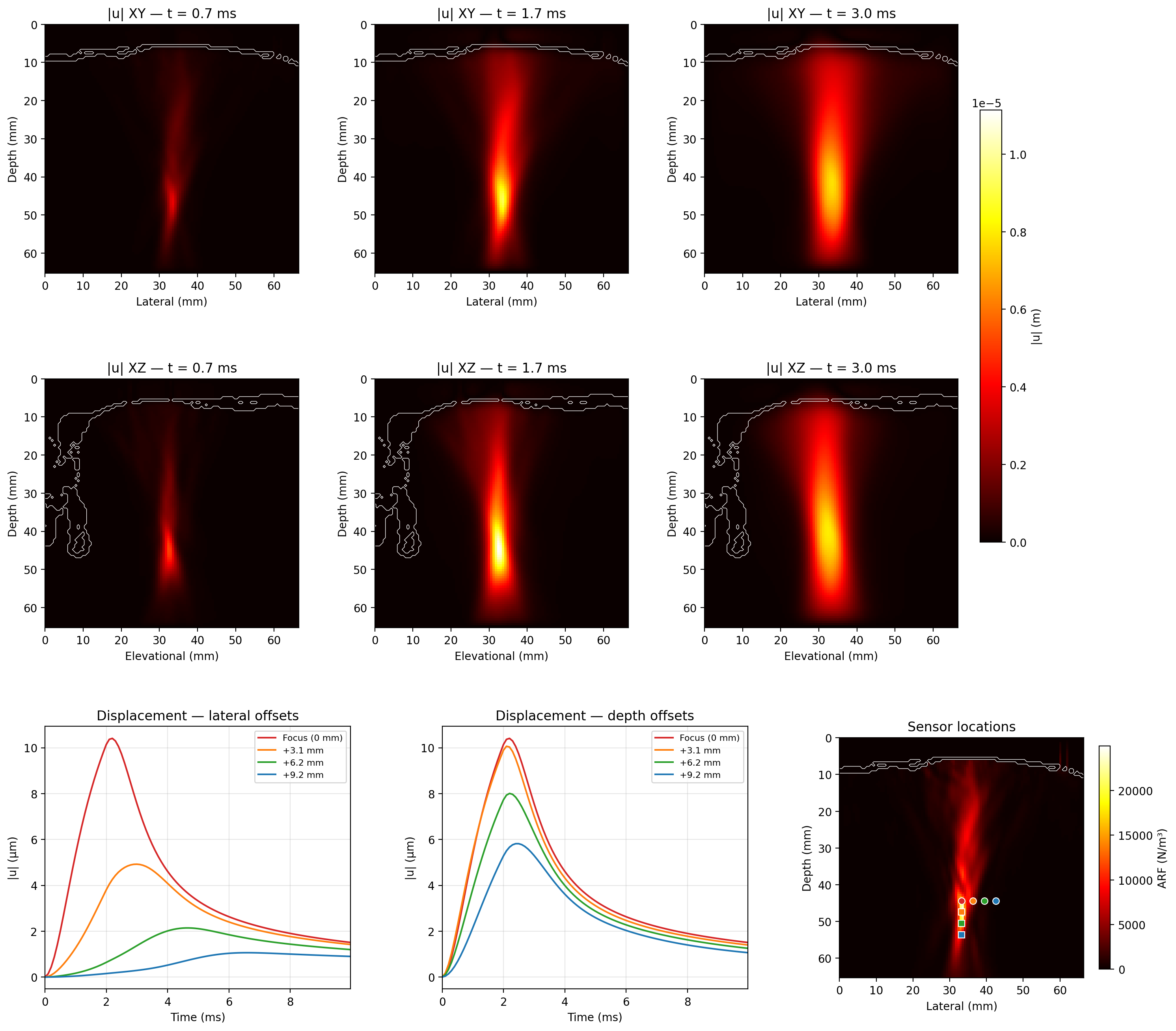}
  \caption{Shear wave displacement for the sparse array configuration.
    \textbf{Top row:} displacement magnitude $|u|$ in $x$--$y$ slices at
    center elevation.
    \textbf{Middle row:} $|u|$ in $x$--$z$ slices at center lateral position.
    White contours indicate the skull boundary.
    \textbf{Bottom row:} displacement time traces at the focal spot and at
    lateral offsets (left) and depth offsets (centre) from the ARF peak.
    Sensor locations are shown on the ARF map (right, circles = lateral,
    squares = depth offsets).}
  \label{fig:displacement_snapshots}
\end{figure}

The shear wave simulation uses the staggered-grid FDTD solver with brain tissue
parameters, shear wave speed $c_s = 2.14$~m/s at 75~Hz, density
$\rho = 1000$~kg/m$^3$, giving shear modulus
$\mu = \rho\,c_s^2 \approx 4580$~Pa. Kelvin--Voigt viscosity is set to
$\eta = 0.5$~Pa$\cdot$s. All three ARF components are applied as forcing terms
via the separable temporal envelope, requiring only a single pre-computed Helmholtz
projection. Peak displacement reaches $|u_x|^{\max} \approx 1.7$~$\mu$m.
Figure~\ref{fig:displacement_snapshots} shows the displacement magnitude field
expanding radially from the focal spot over 0.7--3.0~ms, and displacement time
traces over 10~ms at the focal spot and at lateral and depth offsets of 3.1,
6.2, and 9.2~mm, confirming the expected shear wave propagation speed and
amplitude decay with distance.

\subsection{TIPS curved-bowl transducer through micro-CT skull}
\label{sec:tips_demo}

The second demonstration uses the Philips TIPS (Therapeutic Imaging and
Phased~array Sonication) transducer, a curved-bowl annular phased array
designed for focused ultrasound therapy and neuromodulation. The
TIPS transducer is modeled as a spherical-cap bowl, radius of curvature
$R = 80$~mm, inner aperture radius $r_{\mathrm{in}} = 20.5$~mm, outer aperture
radius $r_{\mathrm{out}} = 46$~mm (92~mm full diameter), 8~concentric annular
elements of equal radial width, and a bowl thickness of 3~voxel layers along
the surface normal.

\begin{figure*}
  \centering
  \includegraphics[width=\textwidth]{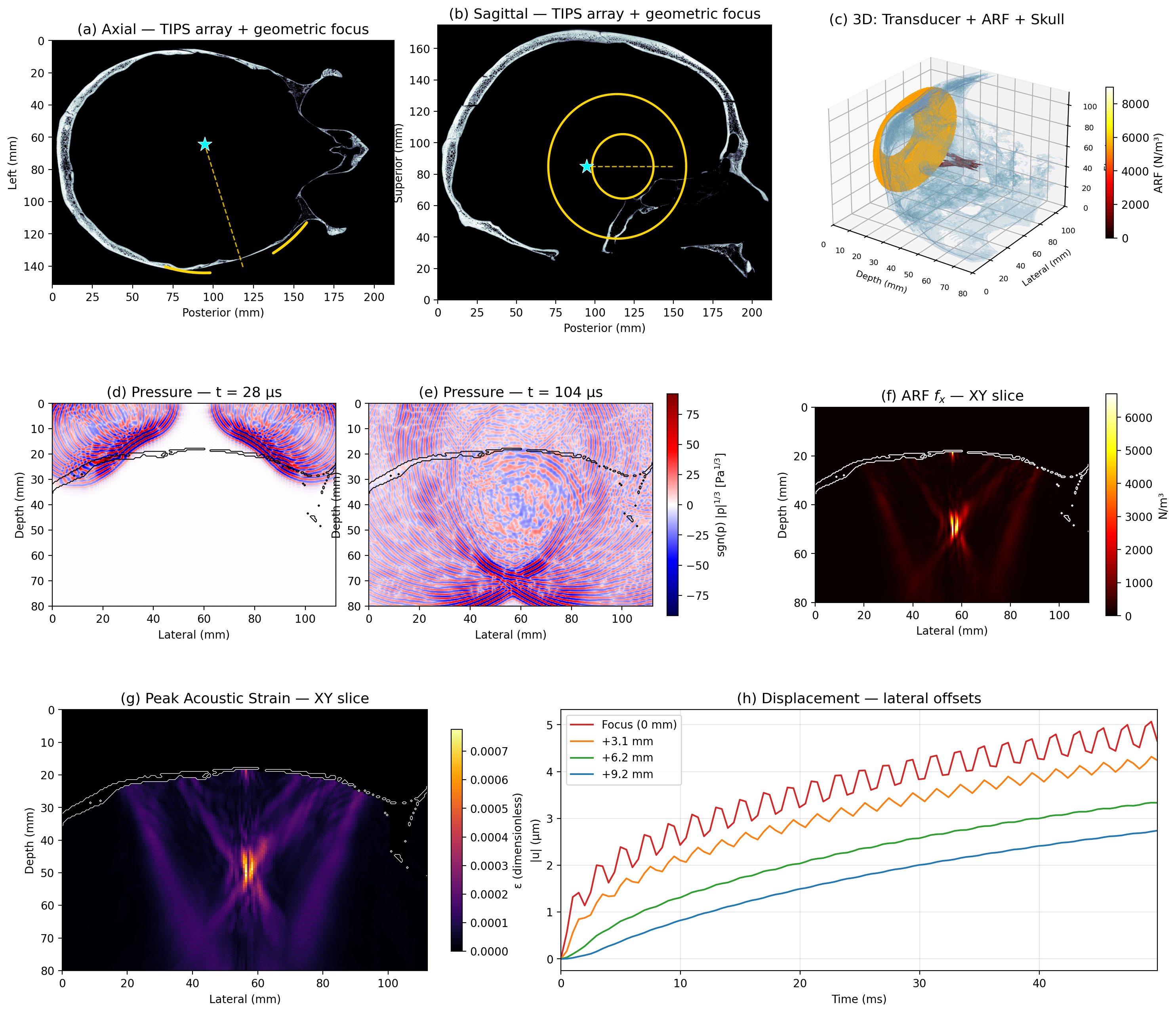}
  \caption{Composite TIPS transcranial shear wave simulation.
    \textbf{(a,b)}~Axial and sagittal micro-CT slices showing the TIPS
    annular aperture (gold rings) and geometric focal spot (cyan star).
    \textbf{(c)}~3D rendering of the skull slab (semi-transparent blue),
    TIPS bowl transducer (gold), and ARF isosurface (coloured by intensity).
    \textbf{(d,e)}~Pressure propagation at $t = 28$ and $104$~$\mu$s
    (skull contour in black).
    \textbf{(f)}~Brain-masked ARF depth component $b_0$ in the $x$--$y$
    plane (skull contour in white).
    \textbf{(g)}~Peak volumetric acoustic strain $\varepsilon$ in the
    $x$--$y$ plane.
    \textbf{(h)}~Displacement magnitude time traces at the focal spot and
    at lateral offsets from the ARF peak.}
  \label{fig:tips_composite}
\end{figure*}

The heterogeneous skull medium uses the same Halle temporal bone micro-CT
specimen as Section~\ref{sec:sparse_demo}. Because the TIPS bowl surface
extends further in depth, the skull slab is shifted deeper to avoid geometric
intersection. The acoustic simulation uses 400~kPa source pressure with
3-cycle Gaussian-modulated sinusoidal pulses at $f_0 = 1$~MHz.
Figure~\ref{fig:tips_composite} summarises the full TIPS transcranial
simulation pipeline.
Panels~(a,b) show the transducer placement on axial and sagittal micro-CT
slices of the Halle skull, with the annular TIPS aperture (gold) and
geometric focal spot (cyan star) overlaid.
Panel~(c) renders the skull slab (semi-transparent), TIPS bowl (gold), and
ARF isosurface (hot colourmap) in 3D.
The geometrically focused bowl produces a converging wavefront that propagates
through the skull and focuses at 50~mm depth.
Figure~\ref{fig:tips_composite}(d,e) shows the pressure field at $t = 28$~$\mu$s
(converging wavefront) and $t = 104$~$\mu$s (post-focal reflections).
The acoustic radiation force is computed using the pressure-based plane-wave
approximation Eq.~\eqref{eq:isppa_norm} with pulse normalization corrections.
The peak body force after brain masking is
$b_0^{\max} \approx 22{,}600$~N/m$^3$.
Figure~\ref{fig:tips_composite}(f) shows the depth component of the radiation
force in the $x$--$y$ plane with brain masking and skull contour overlay.
The shear wave simulation uses the same brain tissue parameters as
Section~\ref{sec:sparse_demo}. The pulsed forcing protocol uses a 50\% duty
cycle (1~ms on, 1~ms off) sustained over 50~ms (25~pulses). Peak displacement
reaches $|u_x|^{\max} \approx 5$~$\mu$m at the focal spot.
Figure~\ref{fig:tips_composite}(h) shows displacement time traces at the
focal spot and at lateral offsets of 3.1, 6.2, and 9.2~mm.
The volumetric acoustic strain $\varepsilon = -p/K$ (where $K = \rho c^2$ is
the bulk modulus) and its spatial gradient are computed directly from the
pressure time series without the pulse-length normalization used for radiation
force. Peak acoustic strain reaches $8.8 \times 10^{-4}$ and peak strain
gradient magnitude reaches $1.5$~m$^{-1}$.
Figure~\ref{fig:tips_composite}(g) shows the peak strain field. These
quantities are relevant to flexoelectric coupling mechanisms proposed for
ultrasonic neuromodulation~\cite{felix2022flexo}.

\section{Discussion}
\label{sec:discussion}

\subsection{Physical assumptions}

The solver is derived under a small-strain linearization, assuming small
displacement gradients and a linear stress--strain relationship. This is
typically appropriate for ultrasound radiation-force excitations in
SWE~\cite{sarvazyan1998elasticity,barr2015elastography}, where observed tissue
motion is commonly in the micrometer to tens-of-micrometers regime.

Material behavior is modeled as isotropic with an effectively homogeneous (or
piecewise homogeneous) shear response, represented by a scalar shear modulus
$\mu(\bm{x})$ and density
$\rho(\bm{x})$~\cite{fung1993biomechanics,duck1990physical}. This
approximation is often reasonable in parenchymal soft tissues at the spatial
resolution used for SWE
processing~\cite{gennisson2013ultrasound}, but it can break down in anisotropic
or strongly heterogeneous structures such as skeletal muscle, tendon,
white-matter tracts, and near rigid interfaces (e.g., skull).

A central modeling choice is to represent shear-dominant propagation while
omitting explicit compressional (longitudinal) elastodynamics. This is
motivated by the near-incompressibility of soft tissue ($K\gg\mu$). The
trade-off is that this reduced model does not capture near-source pressure-wave
content, P--S mode conversion at interfaces, nonlinear shear or shear shock
dynamics~\cite{tripathi2017shearshock,espindola2017shearshock,tripathi2019shearshock2d},
or any diagnostic relying on volumetric strain.

Viscous attenuation is represented using a Kelvin--Voigt shear
viscosity~\cite{catheline2004viscoelastic,ferry1980viscoelasticity}. This
single-parameter form is useful for capturing qualitative attenuation and pulse
broadening, but soft tissue often exhibits power-law or fractional viscoelastic
behavior~\cite{urban2012review}. A constant $\eta$ may be insufficient for
broadband inversions or dispersion-sensitive analyses.

Finally, the present implementation enforces fixed (zero-displacement)
boundaries. Practical mitigation strategies include using larger computational
domains, restricting analysis to earlier times, or implementing absorbing layers
(e.g., PML)~\cite{berenger1994perfect,collino2001application}.


\subsection{Computational cost}

The reduced FDTD formulation is computationally lightweight compared to
general-purpose FEM solvers. For the TIPS transcranial demonstration
(Section~\ref{sec:tips_demo}), the shear wave simulation uses a
$129 \times 181 \times 181$ grid ($4.2 \times 10^6$ points) and 3009 time
steps. With pre-computed Helmholtz projection, the shear FDTD completes in
approximately 4~minutes on a single CPU core (NumPy, no GPU acceleration).
The QIBA benchmark ($240 \times 200 \times 200$ grid, $9.6 \times 10^6$
points) runs in under 10~minutes under the same conditions. By comparison,
FEM approaches to ARF-driven shear wave simulation using commercial codes
such as LS-DYNA typically require hours of computation on comparable 3D
grids~\cite{palmeri2005fem,palmeri2017fem}. Green's function
methods~\cite{yang2018gpu} can be faster for homogeneous media and sparse
observation points, but do not naturally extend to full-field output or strong
spatial heterogeneity. The present solver occupies a middle ground. It provides
full-field output with spatial heterogeneity at a fraction of the cost of FEM,
while remaining a general-purpose grid solver rather than a point-wise
evaluator.

\subsection{Frequency dependence of ARF magnitude}

Because absorption in soft tissue scales approximately linearly with frequency,
the ARF magnitude depends strongly on $f_0$ even at the same focal intensity
(Table~\ref{tab:arf_scaling}). In clinical soft-tissue ARFI at 5--7~MHz, focal intensities of
500--2000~W/cm$^2$ produce body forces on the order of
$10^4$--$10^5$~N/m$^3$, yielding tissue displacements of 1--20~$\mu$m.
For transcranial applications at lower frequencies (0.5--1~MHz), the absorption
per unit path length is roughly an order of magnitude lower, and the skull
attenuates the beam to roughly 30--50\% of its free-field value. The net effect
is that body forces at 1~MHz are typically $10^2$--$10^4$~N/m$^3$, and the
resulting brain tissue displacements are sub-micrometer to a few micrometers.

\begin{table}
\centering
\caption{Frequency dependence of ARF body force at fixed focal intensity
  ($I_{\mathrm{SPPA}} = 1000$~W/cm$^2$,
   $\alpha_0 = 0.5$~dB\,cm$^{-1}$\,MHz$^{-1}$, $c = 1540$~m/s).}
\label{tab:arf_scaling}
\begin{tabular}{ccc}
\hline
$f_0$ (MHz) & $\alpha_{\mathrm{Np}}$ (Np/m) & $b$ (N/m$^3$) \\
\hline
1   &  5.8  & 75\,300  \\
3   & 17.3  & 224\,700 \\
5   & 28.8  & 374\,000 \\
6.7 & 38.6  & 501\,300 \\
\hline
\end{tabular}
\end{table}

\subsection{Applicability and limitations}
The reduced model is most appropriate in regimes with small displacements and
approximately linear
response~\cite{sarvazyan1998elasticity}, in nearly incompressible soft tissue
where SWE biomarkers depend primarily on the shear-wave
speed~\cite{sandrin2003transient,tanter2008quantitative,barr2015elastography},
in moderate propagation ranges where boundary reflections are not dominant, and
in analyses that are effectively narrowband such that a Kelvin--Voigt damping
term provides an adequate
approximation~\cite{catheline2004viscoelastic}.
The assumptions are more likely to fail in strongly anisotropic or
heterogeneous media, in problems where compressional waves must be represented
explicitly, in regimes where nonlinear shear or shear shock formation is
significant~\cite{tripathi2017shearshock,espindola2017shearshock}, in broadband
rheological inversions, in large-domain or long-duration simulations where fixed-wall
reflections can influence arrival times, and in settings where poroelasticity is needed. Absorbing-boundary (PML) support is a natural extension of the present fixed-wall implementation and is planned as future work.

\section{Conclusion}
\label{sec:conclusion}

This manuscript derived a reduced shear elastodynamic formulation and an
explicit FDTD implementation for ultrasound-driven shear wave propagation in
soft tissue. The governing equation is obtained by systematically applying
near-incompressibility, small-strain linearization, Helmholtz decomposition,
and solenoidal force projection to the full Navier equations, yielding a
Kelvin--Voigt shear wave equation that isolates the transverse dynamics relevant
to radiation-force--induced motion.

Validation in homogeneous benchmark media shows accurate recovery of the
imposed shear wavespeed under controlled conditions ($<$0.1\% error at the
finest grid), second-order convergence under grid refinement, and agreement with
analytical Kelvin--Voigt attenuation and phase speed predictions (${\sim}$3\%
and $<$1\%, respectively). An end-to-end benchmark against the RSNA QIBA
elastic phantom standard confirms accurate shear wave speed recovery
(${\sim}$1\% error) across a tenfold range of shear moduli, validating the
complete acoustic--elastic pipeline including radiation force computation.
The pre-computed Helmholtz projection for separable
body forces reduces simulation time by one to two orders of magnitude in
typical ARF-driven scenarios.

The principal limitations are those implied by the model reduction. Compressional
dynamics, mode conversion at interfaces, anisotropy, strong heterogeneity,
broadband frequency-dependent viscoelasticity, and late-time reflection
artifacts are not fully represented. Within its intended scope, the reduced
solver provides a lightweight, purpose-built tool for multiphysics ultrasound
workflows where shear wave generation, propagation, and measurement are the
dominant quantities of
interest~\cite{sarvazyan1998elasticity,bercoff2004supersonic,barr2015elastography,nightingale2002radiation,tanter2008quantitative}.
End-to-end transcranial demonstrations with sparse phased array and
curved-bowl transducer geometries confirm integration with GPU acoustic
simulators and radiation force computation through realistic skull models,
producing shear displacements of 1.7--5~$\mu$m consistent with clinical ARFI
measurements. The computational efficiency of the reduced solver extends its
applicability to inverse problems, iterative methods, and machine-learning
data-generation tasks.



\funding{This work was supported in part by NIH grants R01-EB037345, RF1-NS113285, and R01-EB036295.}

\data{The simulation code and validation scripts are available at \url{https://github.com/pinton-lab/shearwave}.}

\bibliographystyle{unsrt}
\bibliography{general}

\end{document}